\documentclass[conference]{IEEEtran}
\IEEEoverridecommandlockouts
\usepackage{cite}
\usepackage{amsmath,amssymb,amsfonts,bbm,bm}
\usepackage{algorithm}
\usepackage{algorithmic}
\usepackage{graphicx}
\usepackage{float}
\usepackage{subfigure}
\usepackage{textcomp}
\usepackage{xcolor}
\usepackage{float}
\usepackage{paralist}
\usepackage{multirow}
\usepackage{booktabs}
\usepackage[font=small]{caption}
\usepackage{subcaption}
\usepackage{graphicx}
\usepackage{tikz}
\usetikzlibrary{shapes.geometric, arrows, positioning}
\usepackage[nolist]{acronym}
\usepackage{epstopdf}

\DeclareMathOperator*{\argmax}{argmax}
\DeclareMathOperator*{\argmin}{argmin}

\newtheorem{remark}{Remark}

\newtheorem{lemma}{Lemma}

\newcommand{\mW}{\mathbf{W}}
\newcommand{\mI}{\mathbf{I}}
\newcommand{\mH}{\mathbf{H}}
\newcommand{\mA}{\mathbf{A}}

\newcommand{\mG}{\mathbf{G}}
\newcommand{\mL}{\mathbf{L}}

\newcommand{\mQ}{\mathbf{Q}}
\newcommand{\mT}{\mathbf{T}}
\newcommand{\mU}{\mathbf{U}}
\newcommand{\mLambda}{\boldsymbol{\Lambda}}

\newcommand{\ba}{\mathbf{a}}
\newcommand{\bb}{\mathbf{b}}
\newcommand{\bc}{\mathbf{c}}
\newcommand{\bv}{\mathbf{v}}
\newcommand{\bu}{\mathbf{u}}
\newcommand{\bh}{\mathbf{h}}
\newcommand{\bs}{\mathbf{s}}

\newcommand{\bz}{\mathbf{z}}
\newcommand{\bw}{\mathbf{w}}

\newcommand{\bl}{\mathbf{l}}

\newcommand{\bq}{\mathbf{q}}
\newcommand{\bg}{\mathbf{g}}

\newcommand{\bomega}{\boldsymbol{\omega}}

\newcommand{\setC}{\mathbb{C}}
\newcommand{\setR}{\mathbb{R}}
\newcommand{\setS}{\mathbb{S}}

\newcommand{\R}{\mathcal{R}}
\newcommand{\Lag}{\mathcal{L}}

\newcommand{\sinr}{\mathrm{SINR}}

\newcommand{\trp}{\mathsf{T}}
\newcommand{\her}{\mathsf{H}}
\newcommand{\set}[1]{\left\lbrace #1 \right\rbrace}
\newcommand{\tr}[1]{\mathrm{tr}\left\lbrace #1 \right\rbrace}
\newcommand{\brc}[1]{\left( #1 \right)}
\newcommand{\dbc}[1]{\left[ #1 \right]}
\newcommand{\norm}[1]{\left\Vert #1 \right\Vert}
\newcommand{\abs}[1]{\left\vert #1 \right\vert}
\newcommand{\diag}[1]{\mathrm{diag} \left\lbrace #1 \right\rbrace}
\newcommand{\Ex}[1]{\mathbbm{E} \left\lbrace #1 \right\rbrace}

\def\BibTeX{{\rm B\kern-.05em{\sc i\kern-.025em b}\kern-.08em
    T\kern-.1667em\lower.7ex\hbox{E}\kern-.125emX}}

\begin{document}
\begin{acronym}
\acro{ap}[AP]{access point}
\acro{mimo}[MIMO]{multiple-input multiple-output}
\acro{los}[LoS]{line-of-sight}
\acro{noma}[NOMA]{non-orthogonal multiple access}
\acro{irs}[IRS]{reconfigurable intelligent surface}
\acro{snr}[SNR]{signal to noise ratio}
\acro{sinr}[SINR]{signal to interference and noise ratio}
\acro{bcd}[BCD]{block coordinate descent}
\acro{rzf}[RZF]{regularized zero-forcing}
\acro{pas}[PAS]{pinching antenna system }
\acro{fp}[FP]{fractional programming}
\acro{mrt}[MRT]{maximal-ratio transmission}
\acro{zf}[ZF]{zero-forcing}
\end{acronym}

\title{Downlink Beamforming with Pinching-Antenna Assisted MIMO Systems\vspace{-4mm}}
\author{\IEEEauthorblockN{Ali Bereyhi}
\IEEEauthorblockA{University of Toronto\\
ali.bereyhi@utoronto.ca}
\and
\IEEEauthorblockN{Saba Asaad}
\IEEEauthorblockA{York University \\
asaads@yorku.ca}
\and
\IEEEauthorblockN{Chongjun Ouyang}
\IEEEauthorblockA{Queen Mary University\\
c.ouyang@qmul.ac.uk}
\and
\IEEEauthorblockN{Zhiguo Ding}
\IEEEauthorblockA{University of Manchester\\
zhiguo.ding@manchester.ac.uk}
\and
\IEEEauthorblockN{H. Vincent Poor}
\IEEEauthorblockA{Princeton University\\
poor@princeton.edu}
}

\maketitle

\begin{abstract}
Pinching antennas have been recently proposed as a promising flexible-antenna technology, which can be implemented by attaching low-cost pinching elements to dielectric waveguides. This work explores the potential of employing pinching antenna systems (PASs) for downlink transmission in a multiuser MIMO setting. We consider the problem of hybrid beamforming, where the digital precoder at the access point and the activated locations of the pinching elements are jointly optimized to maximize the achievable weighted sum-rate. Invoking fractional programming, a novel low-complexity algorithm is developed to iteratively update the precoding matrix and the locations of the pinching antennas. We validate the proposed scheme through extensive numerical experiments. Our investigations demonstrate that using PAS the system throughput can be significantly boosted as compared with the conventional fixed-location antenna systems, enlightening the potential of PAS as an enabling candidate for next-generation wireless networks. 
\end{abstract}


\section{Introduction}
\label{sec:intro}
Shannon’s theory established the cornerstone of information theory and has significantly influenced the evolution of modern communication systems \cite{shannon1948mathematical}. The conventional information-theoretic formulation of a communication system considers the channel between the transmitter and receiver to be determined by environmental factors, and hence to be beyond human control and manipulation. 
However, recent advances in communication technologies have challenged this assumption by introducing techniques with the capabilities to reconfigure channel parameters. An early example is the introduction of \ac{mimo} systems, which can modify the end-to-end channel by establishing parallel links between a pair of transmitter and receiver \cite{lu2014overview}.

Recent developments in the area of flexible antenna design have provided more sophisticated means of channel reconfiguration. Technologies, such as \acp{irs} \cite{bereyhi2023channel}, fluid antennas \cite{wong2020fluid}, and movable antennas \cite{zhu2023movable}, have provided significant degrees of freedom to control the effective end-to-end channel gain through their reconfigurable parameters. 
For example, \acp{irs} reshape the wireless propagation environment in favor of signal transmission, such as bypassing obstructions, improving signal coverage, and reducing multi-user interference \cite{zheng2021double}. Another example is fluid antenna technology, which provides reconfigurability by employing a metallic or non-metallic liquid, e.g., Mercury or Galinstan, within a dielectric holder for signal transmission. This enables real-time adjustments to the shape and position of the antennas, making them suitable for compact devices with limited space \cite{wong2020fluid}. The movable antenna technology enables the localized movement of antenna elements in the transmitter/receiver array within designated regions, effectively using spatial degrees of freedom to improve wireless channel conditions in dynamic environments \cite{zhu2023movable}.

\subsection{Pinching-Antenna Systems}
Despite their promising theoretical gains in terms of system throughput, the integration aspect of the mentioned technologies remains challenging due to the need for real-time control, high cost and complexity, and limited ability to address large-scale path loss. Considering these challenges, several lines of research have focused on developing alternative reconfigurable technologies. The \ac{pas} is one of the most recent proposals, which was originally proposed and demonstrated by NTT DOCOMO in 2022; see \cite{suzuki2022pinching,ding2024flexible,yang2025pinching}. This innovative approach leverages low-cost dielectric materials, such as plastic clothespins, which can be applied at any point along a dielectric waveguide for radiation. 

\acp{pas} offer flexibility in establishing or strengthening \ac{los} links by activating pinching elements close to the serving user \cite{ding2024flexible}. Unlike many earlier proposals, the tunable elements of \acp{pas}, i.e., pinching elements, can be implemented in a straightforward and low-cost manner. This has attracted several lines of recent research in the literature. The ergodic rate achieved by \ac{pas} was analyzed in the primary work \cite{ding2024flexible} using stochastic geometry tools. The array gain achieved by \ac{pas} was examined in \cite{ouyang2025array}. Several activation algorithms for optimizing the locations of pinching elements along a single waveguide were proposed in \cite{xu2024rate,wang2024antenna,tegos2024minimum}. These studies collectively highlight the superiority of \ac{pas} over conventional fixed-location antenna systems in enhancing system throughput.

\subsection{Contributions}
The initial study in \cite{ding2024flexible} has investigated the potential of \ac{pas} for \ac{mimo} communication, considering a simple two-user scenario. Motivated by their results, this work aims to study a gap that has not been explored in the earlier studies: the potential of \ac{pas} for general multiuser downlink transmission. To this end, we study the problem of multiuser beamforming through the following lines of contribution:
\begin{inparaenum}
\item[($i$)] we formulate downlink transmission in a multiuser setting with multiple pinched waveguides as a classic vector Gaussian broadcast channel with a reconfigurable channel matrix.
\item[($ii$)] We propose a \textit{hybrid} beamforming design in which the locations of the pinching elements and the digital precoding matrix are designed jointly, such that the weighted achievable sum-rate is maximized.
\item[($iii$)] We develop a tractable iterative algorithm for the proposed joint design, which optimizes the digital precoding matrix at the transmitter and adjusts the locations of the pinching elements, simultaneously. 
\end{inparaenum}
Numerical simulations demonstrate that the proposed algorithm enables \ac{pas}-aided transmitter to achieve a significantly higher sum-rate, as compared with conventional fixed-location antenna systems, which highlights the great potential of this technology for improving the throughput of next-generation wireless systems. 

\paragraph*{Notation} Vectors, and matrices are shown by bold lower-case and bold upper-case letters, respectively. The transpose, conjugate and conjugate-transpose of $\mH$ are denoted by $\mH^{\trp}$, $\mH^*$ and $\mH^{\her}$, respectively. The $N\times N$ identity matrix is shown by $\mI_N$, and $\setR$ and $\setC$ denote the real axis and complex plane, respectively. For set $\{1,\ldots,N\}$, we use shortened notation $[N]$.  We denote expectation by $\Ex{.}{}$, and $\mathcal{CN}\brc{\eta,\sigma^2}$ is the complex Gaussian distribution with mean $\eta$ and variance $\sigma^2$. When the summation range is clear from the context, we omit it and denote only the index, e.g., $\sum_{k}$. 

\section{System Model and Problem Formulation}
\label{sec:formulation}
\begin{figure}[!t]
\centering
\setlength{\abovecaptionskip}{0pt}
\includegraphics[height=0.15\textwidth]{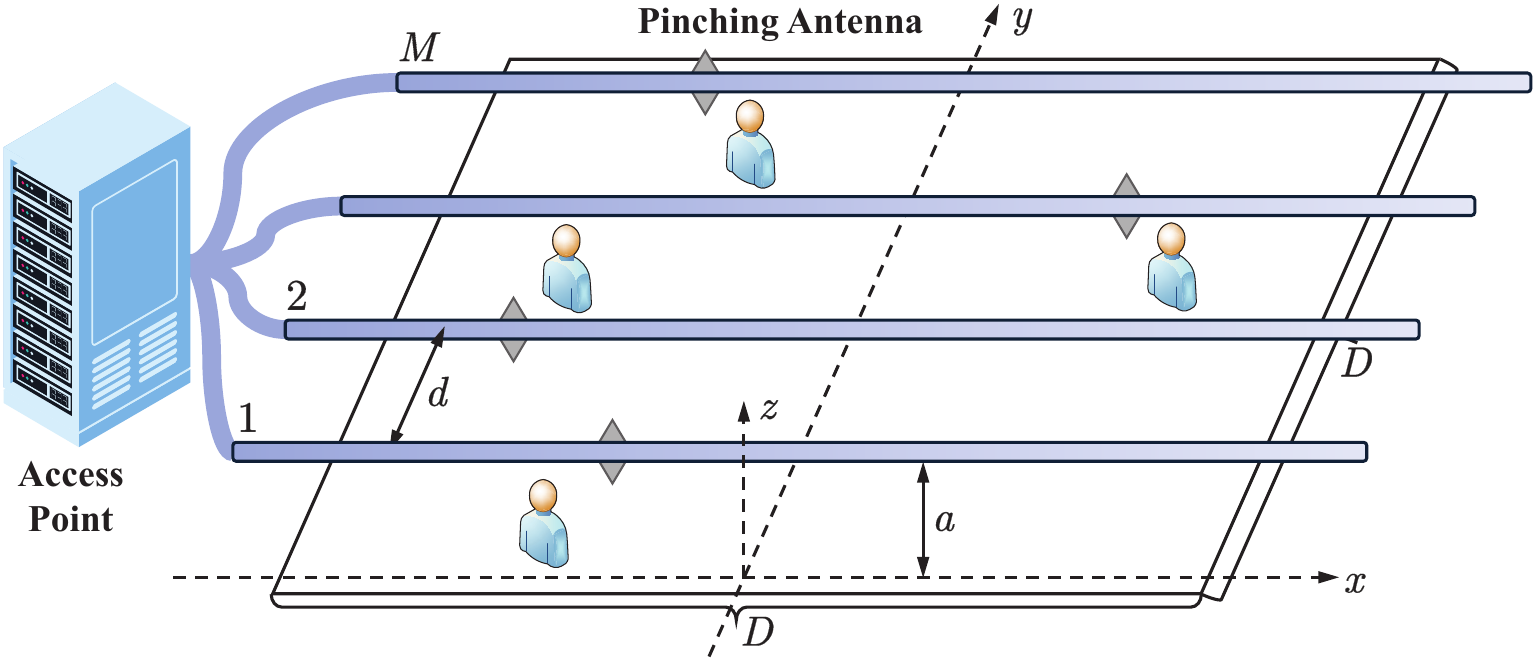}
\caption{Illustration of the system configuration: AP is equipped by multiple waveguides each pinched with an element.}\vspace{-6mm}
\label{Fig1_Schematic}
\end{figure}

Consider $M$ dielectric waveguides, each equipped with a pinching element that can freely move across the waveguide. In practice, this can be realized via multiple elements, each covering one part of the waveguide. The element on each waveguide acts as an isotropic radiator whose signal is the phase-shifted version of the signal fed~to the waveguide. Fig.~\ref{Fig1_Schematic} shows the configuration. We assume that waveguides are extended over the $x$-axis at the altitude $a$ in an array formed on the $y$-axis with each two waveguides being distanced $d$. The location of the pinching element on the waveguide $m\in [M]$ is $\bv_{m} = [\ell_{m}, (m-1)d, a]$, where $\ell_{m}$ is the position of the element on the waveguide $m$, assuming that the signal is fed to the waveguide at $[0, (m-1)d, a]$. Note that $\ell_{m}$ for $m\in[M]$ are the design parameters to be optimized.

The waveguides are fed by an \ac{ap} that aims to serve $K$ single-antenna users. The users are distributed within a known two-dimensional area, e.g., a square region. We denote the location of user $k\in [K]$ by $\bu_k = [x_k, y_k, 0]$, assuming they are located in the $xy$-plane.

Let $z_m$ denote the signal fed to waveguide $m$. The pinching element on this waveguide radiates a phase-shifted version of $z_m$. The radiated signal from the pinching elements can hence be represented as
$t_m = \exp\set{-j \theta_{m}} z_m$, where $\theta_m$ denotes the phase-shift at the pinching element on the waveguide $m$ and is determined by the location of the element. Noting that $z_m$ is fed at location $\ell_{m} = 0$, at the carrier frequency $f$ we can write $\theta_{m} =  {2 \pi i_{\mathrm{ref}}} \abs{\ell_{m}}/ {\lambda}$, where $\lambda = \mathrm{c}/f$ denotes the wavelength with $\mathrm{c}$ being the speed of light, and $i_{\mathrm{ref}}$ is the reflective index of the waveguides. This indicates that the radiated signal is a function of $\ell_m$, i.e., $z_m(\ell_m)$. We however drop this dependency for ease of notation.

\subsection{Characterizing End-to-End Channel}
Similar to \cite{ding2024flexible}, we assume that the users are in the \ac{los} of the waveguides, which is typically the case in indoor settings. Furthermore, those non-\ac{los} paths are assumed to be negligible since they are typically much weaker than those LoS paths. Let $y_k$ denote the signal received by the user $k$. Considering \ac{los} channels, we can write $y_k$ as 
\begin{align}
    y_k = \sum_{m} h_{m,k} t_m + \varepsilon_k,
\end{align}
where $\varepsilon_k$ is complex zero-mean Gaussian noise with variance $\sigma^2$, i.e., $\varepsilon_k\sim\mathcal{CN} \brc{0,\sigma^2}$, and $h_{m,k}\in \setC$ denotes the channel coefficient from the element on waveguide $m$ to user $k$.

The channel coefficient $h_{m,k}$ can be explicitly expressed in terms of the location of the user and pinching element as \cite{ding2024flexible}
\begin{align}
    h_{m,k} =
    \frac{\xi \alpha_{m,k} }{\norm{\bv_{m} - \bu_k}} 
    \exp\set{-j k_0 \norm{\bv_{m} - \bu_k} },
\end{align}
where $\xi = {\lambda}/{4 \pi}$ is the coefficient proportional to the effective surface of the element, 
$k_0={2\pi}/{\lambda}$ denotes the wavenumber, and $\alpha_{m,k}$ is the coefficient capturing shadowing between the pinching element on waveguide $m$ and user $k$. 

We note that the end-to-end channel between the waveguide array and the user $k$ depends on the element locations. Defining $\bl = [\ell_1, \ldots, \ell_M]^\trp$, this means that the channel is a function of $\bl$. Considering the radiation vector of waveguide $m$, we can write the channel to the user $k$ as
\begin{align}
    y_k = \sum_{m} g_{m,k} \brc{\bl} z_m + \varepsilon_k = {\bg}_{k}^\trp \brc{\bl} \bz + \varepsilon_k,
\end{align}
where $\bz = \dbc{z_{1}, \ldots, z_M}$ denotes the transmitted beam by the array of waveguides, and $\bg_{k} \brc{\bl} = \dbc{{g}_{1,k} \brc{\ell_1}, \ldots, {g}_{M,k} \brc{\ell_M} }^\trp$ with ${g}_{m,k} \brc{\ell_m}$ being the effective channel between the waveguide $m$ and the user $k$ given by
\begin{subequations}
\begin{align}
   {g}_{m,k} \brc{\ell_m} &= h_{m,k}  \exp\set{-j \theta_{m}} \\
    &=
    \frac{ \xi \alpha_{m,k} \exp\set{-j k_0 \brc{D_{m,k}\brc{\ell_m} \hspace{-.5mm}+\hspace{-.5mm} i_{\mathrm{ref}} \ell_{m} } }}{ D_{m,k}\brc{\ell_m} }.
\end{align}
\end{subequations}
Here, $D_{m,k}\brc{\ell_m}$ denotes the distance between the pinching element on the $m$-th waveguide and the user $k$, i.e.,
\begin{align}
D_{m,k}^2\brc{\ell_m} 
=(\ell_m-x_k)^2 + \brc{(m-1)d - y_k}^2 + a^2.
\end{align}
\begin{remark}
The effective channel between the \ac{ap} and user $k$ can be decomposed as
\begin{align}
\bg_k \brc{\bl} = \mL_k \brc{\bl} \bg_k^0
\end{align}
where $\bg_k^0 = \xi [\alpha_{1,k}, \ldots, \alpha_{M,k}]^\trp \in \setC^M$ purely depends on the propagation environment and is independent of the element locations $\ell_1, \ldots, \ell_M$. The matrix $\mL_k \brc{\bl}$ is further given by
\begin{align}
    &\mL_k \brc{\bl} \hspace{-.5mm} =\hspace{-.5mm}  \diag{
    \frac{ \exp\set{-j k_0 \brc{D_{m,k}\brc{\ell_m} \hspace{-.5mm} +\hspace{-.5mm}  i_{\mathrm{ref}} \ell_{m} } }}{ D_{m,k}\brc{\ell_m} } 
    }_{m=1}^M
\end{align}
which depends on the location of pinching elements. 
Note that the channel in this case resembles the same model as in \ac{irs}-assisted channels with a key difference: the channel of each user is modified by a phase-shift and attenuation that is \textit{specific to the user}, i.e., $\mL_k\brc{\bl}$ depends on $k$. This difference comes from the fact that unlike reflecting surfaces, pinching elements can move a large range of distances.
\end{remark}

\subsection{Downlink Beamforming}
Each waveguide is fed by a \textit{single} signal, and the pinching elements can only modify the phase-shift by changing their location. When used in the form of an array, we can achieve multiplexing gain by performing transmit beamforming. Let $\bs = [s_1, \ldots,s_K]^\trp$ denote the collection of transmit signals the \ac{ap} intends to send, with $s_k\in \setC$ being the encoded information signal of user $k$. We assume that the signals are zero-mean unit-variance stationary processes satisfying
    $\Ex{\bs\bs^\her} = \mI$. 
The \ac{ap} sets its feeding signal to each waveguide to be a linear combination of these signals, i.e., it sets $z_m = \Tilde{\bw}_m^\trp \bs$ for some vector of coefficients $\Tilde{\bw}_m \in \setC^K$ specified for waveguide $m$. The signal transmitted by the waveguide array can be represented as $\bz = \mW \bs$ with $\mW = \dbc{\Tilde{\bw}_1, \ldots, \Tilde{\bw}_M}^\trp$ being the beamforming matrix. We further consider a power constraint on $\bz$ as $\mathbbm{E}\{\norm{\bz}^2\} \leq P$ for some $P>0$, which reduces to 
${\rm{tr}}\{\mW \mW^\her\} \leq P$. The signal received by user $k$ is  given by
\begin{align}
    y_k &= \bg_{k}^\trp \brc{\bl} \mW \bs + \varepsilon_k ,
\end{align}
which is a function of both $\mW$ and $\bl$. 


\section{Hybrid Beamforming Design}
In this setting, we can design the \textit{digital} precoder $\mW$ and the \textit{analog} parameter $\bl$, jointly. This is a hybrid beamforming problem, which we discuss in the sequel.

To proceed with the beamforming design, let us first rewrite the received signal at user $k$ as
\begin{align}
    y_k = \sum_{j} \bg_{k}^\trp\brc{\bl} \bw_j s_j + \varepsilon_k,
\end{align}
with $\bw_j$ denoting the $j$-th column vector of $\mW$ collecting the $j$-th entries of $\Tilde{\bw}_m$ for $m\in [M]$. We refer to $\bw_j$ as the \textit{digital beamforming vector} of user $j$. The achievable rate of the user $k$ in this case can be written as
\begin{align}
    R_k \brc{\mW, \bl} = \log \brc{1 + \sinr_k \brc{\mW, \bl}},
\end{align}
where $\sinr_k\brc{\mW, \bl}$ denotes the \ac{sinr} at user $k$ given by
\begin{align}
    \sinr_k \brc{\mW, \bl} = \frac{\abs{\bg_k^\trp  \brc{\bl} \bw_k}^2}{\sum_{j\neq k} \abs{\bg_k^\trp \brc{\bl} \bw_j}^2 + \sigma^2 }.
    \label{eq:SINR_k}
\end{align}
The weighted sum-rate is then given by
\begin{align}
    \R \brc{\mW, \bl} = \sum_{k} \lambda_k R_k \brc{\mW, \bl},
\end{align}
for some weighting coefficient $\lambda_k$ that are proportional to the expected quality of service for user $k$. 

\subsection{Optimal Beamforming Design}
The ultimate goal of \ac{ap} is to maximize the weighted sum-rate for a given power budget $P$ and the physical limitations of the waveguides. This can be formulated as 
\begin{align} \label{eq:optim}
    \hspace{-.05cm}\max_{\mW, \bl} \R \brc{\mW, \bl} 
    ~~~ \mathrm{s.t.} \; \tr{\mW^\her \mW} \leq P,~0\leq\ell_{m}\leq L_m,
\end{align}
where $L_m$ denotes the length of the waveguide $m$. 

The optimization in \eqref{eq:optim} is a non-convex problem whose global solution cannot be computed feasibly and is approximated with a local solution. We hence aim to develop an efficient suboptimal algorithm for this problem. To this end, we start with transforming the problem \eqref{eq:optim} into a \textit{variational} form without the power constraint using the following lemmas.
\begin{lemma}\label{Equality Power Constraint}
The optimal solution to the problem \eqref{eq:optim} satisfies the power constraint with equality, i.e., $\tr{\mW_{\star}^{\her} \mW_{\star} } = P$, for any $\mW_\star$ that is a solution to \eqref{eq:optim}.
\end{lemma}
\begin{IEEEproof}
The proof is concluded by contradiction: let $\hat{\mathbf{W}}=[{\hat{\mathbf{w}}}_1,\ldots,{\hat{\mathbf{w}}}_K]$ be a solution to problem \eqref{eq:optim} that satisfies
\begin{align}
\hat{P}\triangleq \tr{\hat{\mathbf{W}}^\her\hat{\mathbf{W}}}=\sum_{j} \norm{{\hat{\mathbf{w}}}_j}^2<P.    
\end{align}
We now define the scaling factor $\rho = P/\hat{P}$ and the a scaled solution $\bar{{\mathbf{w}}}_j=\sqrt{\rho}{\hat{\mathbf{w}}}_j$. Note that the scaled beamforming matrix $\bar{\mW}= [{\bar{\mathbf{w}}}_1,\ldots,{\bar{\mathbf{w}}}_K]$ satisfies the power constraint with equality. 
Replacing into \eqref{eq:SINR_k}, it is readily shown that 
\begin{align}
    \sinr_k \brc{\bar{\mW}, \bl} > \sinr_k (\hat{\mW}, \bl),
\end{align}
for any $\bl$. This means that $\bar{\mW}$ achieves a larger sum-rate, which contradicts with the initial assumption. 
\end{IEEEproof}

\begin{lemma}\label{lem:2}
For a given $\bl$, let $\bar{\mW}= [{\bar{\mathbf{w}}}_1,\ldots,{\bar{\mathbf{w}}}_K]$ denote a  solution to the following problem 
\begin{align}\label{Beamforming_Problem_2}
\max_{\mW}  \sum_{k} \lambda_k \log(1+\overline{\sinr}_k \brc{\mW,\bl}),
\end{align}
where $\overline{\sinr}_k\brc{\mW,\bl}$ for ${\mW}= [{{\mathbf{w}}}_1,\ldots,{{\mathbf{w}}}_K]$ is defined as
\begin{align}\label{eq:sinr_k_bar}
\overline{\rm{SINR}}_k \brc{\mW,\bl} =\frac{\abs{\bg_k^\trp\brc{\bl}  \bw_k}^2}{\sum_{j\neq k} \abs{\bg_k^\trp\brc{\bl}  \bw_j}^2 + \tfrac{\sigma^2}{P}\sum_{j}\norm{{{\mathbf{w}}}_j}^2 }.
\end{align}
Then, a solution to \eqref{eq:optim} is determined from $\bar{\mW}$ as 
\begin{align}\label{Scaled_Solution}
\breve{\mW} =\sqrt{{P}/{\tr{\bar{\mW}^{\her} \bar{\mW}}}} \bar{\mW}.
\end{align}
\end{lemma}
\begin{IEEEproof}
It is readily shown that the solution in \eqref{Scaled_Solution} satisfies the power constraint with equality and that 
\begin{align}
\overline{\rm{SINR}}_k \brc{\breve{\mW},\bl} = {\rm{SINR}}_k \brc{\breve{\mW},\bl},
\end{align}
for any $\bl$, which implies that the objective of \eqref{eq:optim} attains the same value as the one in \eqref{Beamforming_Problem_2}. Noting that $\bar{\mW}$ is a solution to \eqref{Beamforming_Problem_2}, we can conclude that $\breve{\mW}$ maximizes the objective in \eqref{eq:optim} among all beamformers that satisfy the power constraint with equality. Lemma \ref{Equality Power Constraint} further indicates that no solution with inequality constraint exists. This completes the proof. 
\end{IEEEproof}

Using Lemma~\ref{lem:2}, the optimal hybrid beamforming is equivalently determined by solving the following optimization
\begin{align}\label{Beamforming_Problem_simple}
    \max_{\mW, \bl} \bar{\R} \brc{\mW, \bl}
    \qquad &\mathrm{s.t.} \; 0\leq\ell_{m}\leq L_m,
\end{align}
where $\bar{\R} \brc{\mW, \bl}$ is given by
\begin{align}
\bar{\R} \brc{\mW, \bl}&=\sum_{k} \lambda_k \log \brc{1+\overline{\sinr}_k\brc{\mW, \bl}}
\end{align}
with $\overline{\sinr}_k\brc{\mW, \bl}$ being defined in \eqref{eq:sinr_k_bar}.

\subsection{Variational Solution via Fractional Programming}
The variational problem in \eqref{Beamforming_Problem_simple} describes  maximizing sum of log ratios, whose solution can be efficiently approximated via \ac{fp} as outlined next \cite{shen2018fractional,shen2018fractional2}. We start by defining $\setS$ to be the feasible set of the variational problem \eqref{Beamforming_Problem_simple}. 
Denoting $\bs = \brc{\mW, \bl} \in \setS$, the Lagrange dual transform of \eqref{Beamforming_Problem_simple} is given by 
\begin{align}
\Lag \brc{\bs, \bomega} = \sum_{k} \lambda_k 
\brc{
\log\brc{1+\omega_k} - \omega_k +
\gamma\brc{\bs, \bomega}
},
\end{align}
where $\gamma\brc{\bs, \bomega} = \brc{1+\omega_k} \abs{\bg_k^\trp \brc{\bl} \bw_k}^2 / \Gamma \brc{\bs}$ with 
\begin{align}
\Gamma\brc{\bs}\triangleq{ \sum_j\abs{\bg_k^\trp\brc{\bl} \bw_j}^2 
+ \frac{\sigma^2}{P}\sum_{j}\norm{{{\mathbf{w}}}_j}^2
}.
\end{align}

The Lagrange duality implies that $\bs^\star$ defined as
\begin{align}
    \brc{\bs^\star, \bomega^\star} = \argmax\nolimits_{\bs \in \setS, \bomega \in \setR_+^K} \Lag \brc{\bs, \bomega}
\end{align}
recovers the solution of \eqref{Beamforming_Problem_simple}, and that the optimal values of objectives in both problems are identical. Although the dual problem is not convex, its solution can be efficiently approximated via the \ac{bcd} algorithm \cite{shen2018fractional,shen2018fractional2}. To this end, we initiate $\bs = \brc{\mW, \bl}$ and iterate between the following two marginal problems:
\begin{enumerate}
    \item For fixed $\bs = \brc{\mW, \bl}$, we solve the marginal optimization
    \begin{align}
     \bomega^\star = \argmax\nolimits_{\bomega \in \setR_+^K} \Lag \brc{\bs, \bomega},
\end{align}
whose solution is given by 
\begin{align}\label{fp_bcd_omage_k}
  \omega_k^\star = \overline{\sinr}_k\brc{\mW, \bl}.  
\end{align}
\item We set $\omega_k = \omega_k^\star$ and solve
 \begin{align}\label{eq:Fractional}
     \bs^\star &= \argmax\nolimits_{\bs \in \setS} \Lag \brc{\bs, \bomega^\star},
\end{align}
which reduces to the problem of maximizing~sum~of~fractions. This problem can be represented in quadratic form using the \textit{quadratic transform} \cite{shen2018fractional,shen2018fractional2}.
\end{enumerate}

To tackle the optimization in \eqref{eq:Fractional}, we use the quadratic dual transform and rewrite the optimization as
 \begin{align}
   &  \max_{\bs \in \setS, \bq \in \setC^K} \hspace{-1mm}
\sum_{k} \hspace{-.5mm}\lambda_k\hspace{-1mm}
\left(\hspace{-.5mm} 2 \hspace{-.5mm}\sqrt{1 \hspace{-.5mm} + \hspace{-.5mm} \omega_k^\star}\Re \set{ q^*_k \bg_k^\trp \hspace{-.5mm}\brc{\bl} \hspace{-.5mm}\bw_k } \hspace{-.5mm}
- \hspace{-.5mm} \abs{q_k}^2 \hspace{-.5mm} \Gamma\hspace{-.5mm}\brc{\bs} \hspace{-.5mm}\right)
\label{eq:Quadratic}
\end{align}
This dual problem has a quadratic objective with a smoother landscape and is optimized more efficiently. Note that the problem is still non-convex through the auxiliary variable $\bq$, multiplicative expressions, and the functional form of $\bg_k\brc{\bl}$. We hence develop a two-tier  \ac{bcd} algorithm to approximate the solution of this quadratic problem.

\subsection{Two-tier Iterative Algorithm}
For brevity, we denote the multiuser downlink channel with $\mG \brc{\bl} = \left[\bg_1 \brc{\bl}, \ldots,\bg_K \brc{\bl} \right]^\trp\in \setC^{K\times M}$. We start with the outer loop, in which we use \ac{bcd} to iteratively approximate $\bq^\star$. The outer tier iterates between these two steps:
\begin{enumerate}
    \item Fix $\bs = \brc{\mW, \bl}$ and solve \eqref{eq:Quadratic} for $\bq$. The solution to this  problem is readily given by
    \begin{align}\label{fp_bcd_q_k}
           q_k^{\star} \!=\! \frac{ \sqrt{1+\omega_k^\star}\bg_k^\trp \brc{\bl}  \bw_k }{
           \Gamma\brc{\bs}}.
    \end{align}
    \item Set $\bq = \bq^\star$, and update $\bs$ by solving \eqref{eq:Quadratic} for $\bs$, i.e., 
    \begin{align}\label{eq:lW_problem_Final}
        \hspace{-.2cm} \max_{\mW, \bl} 
         F\brc{\mW, \bl}
        \qquad\hspace{-.2cm} 
        &\mathrm{with} \; 0 \!\leq \!\ell_{m}\leq L_m, 
    \end{align}
    where the objective function, after a few lines of standard derivation, is represented as in \eqref{Obj} at the top of the next page.
    \begin{figure*}[t!]
    \begin{align}\label{Obj}
        &F\brc{\mW, \bl} = 2 \Re\set{ \tr{\mT^\her \mG \brc{\bl} \mW} }- \tr{ \mG \brc{\bl} \mW\mW^\her \mG^\her \brc{\bl} \mU  } 
        - \frac{\sigma^2\tr{\mU}}{P}\tr{\mW\mW^\her},
    \end{align}
   \hrule
    \end{figure*}
In \eqref{Obj}, $\mT = \mQ\mA \mLambda $ and $\mU = \mQ \mLambda \mQ^\her$ for diagonal matrices $\mLambda = \diag{\lambda_k}_{k=1}^K$, $\mQ=\diag{q_k^\star}_{k=1}^K$, and $\mA = \diag{\sqrt{1+\omega_k^\star}}_{k=1}^K$. 
\end{enumerate}

The second marginal problem is non-convex, due to the multiplicative term $\mG\brc{\bl}\mW$. The problem is however marginally convex in both $\mW$ and $\mG\brc{\bl}$. We hence use the \ac{bcd} algorithm once again to solve this marginal problem. The inner loop hence iterates between the following two steps:
\begin{enumerate}
    \item Fix $\bl$, and solve \eqref{eq:lW_problem_Final} for $\mW$, which reduces to \eqref{RZF2} given at the top of next page. This is standard \ac{rzf} precoding,
        \begin{figure*}[t!]
        \begin{equation}\label{RZF2}
        {\mathbf{W}}^{\star}=\argmax_{\mW\in{\mathbb{C}}^{M\times K}} 
        2 \Re\set{ \tr{\mT^\her \mG \brc{\bl} \mW} } - \tr{ \!\mW\mW^\her \!\brc{\!\mG^\her \brc{\bl} \mU \mG\brc{\bl}
        +
        \frac{\sigma^2\tr{\mU}}{P}\mI_M\!}\! },
    \end{equation}
    \hrule
    \end{figure*}
whose solution is given by
            \begin{align}
        {\mathbf{W}}^{\star}
        \!=\!\brc{\!\mG^\her\! \brc{\bl} \mU \mG \brc{\bl}\!+\!\frac{\sigma^2\tr{\mU}}{P}\mI_M\!}^{-1}\!\!\!\!\mG^\her\!\brc{\bl}\mT,\label{fp_bcd_W}
    \end{align}
    and can be seen as \ac{rzf} with an effective channel. 
    
    \item We next set $\mW=\mW^\star$ and solve \eqref{eq:lW_problem_Final} marginally for $\bl$. 
    \end{enumerate}

In the second marginal optimization, the objective is given in terms of exponential sums, and hence its global optimum is not feasibly found. The classical approach to tackle these problems is to invoke greedy approaches. In the sequel,~we develop a greedy approach based on the Gauss-Seidel scheme to efficiently approximate the solution of this marginal problem.



The Gauss-Seidel-based approach suggests to sequentially update the locations with each $\ell_m$ being updated individually treating the other locations as constants. In particular, by fixing $\ell_{m'}$ for ${m'\ne m}$, the scalar problem for $\ell_m$ is given by \eqref{Optimization_single_location} at the top of next page,
\begin{figure*}[t!]
    \begin{align}\label{Optimization_single_location}
        &\max_{ \ell_m} 2 \Re\set{\tr{\mT^\her \tilde{\mG} \brc{\ell_m} \mW} } - \tr{ \mW\mW^\her \tilde{\mG}^\her \brc{\ell_m}\mU \tilde{\mG} \brc{\ell_m} } \;\mathrm{s.t.} \; 0\leq\ell_{m}\leq L_m,
    \end{align}
    \hrule
    \end{figure*}
    where $\tilde{\mG} \brc{\ell_m}$ is defined as  
    \begin{align}
    \tilde{\mG} \brc{\ell_m}=[\tilde{\bg}_1,\ldots,\tilde{\bg}_{m-1},\tilde{\bg}_m\brc{\ell_m},\tilde{\bg}_{m+1},\ldots,\tilde{\bg}_{M}],
    \end{align}
    with $\tilde{\bg}_{m} \brc{\ell_m}\in{\mathbb{C}}^{K\times1}$ representing the $m$-th column of $\mG\brc{\bl}$ whose $k$-th entry is given by
    \begin{align}\label{definition_h_l_m}
    \dbc{\tilde{\bg}\!\brc{\ell_m}}_k =
    \frac{ \xi \alpha_{m,k} \exp\set{-j k_0 
    \brc{D_{m,k}\brc{\ell_m} + i_{\mathrm{ref}} \ell_{m} }
    }}{ D_{m,k}\brc{\ell_m}} .
    %
    \end{align}
Noting that $\ell_{m'}$ for ${m'\ne m}$ is treated as fixed, the argument $\ell_{m'}$ in $\tilde{\bg}_{m'}$ for ${m'\ne m}$ is dropped for clarity. By simple lines of derivation, the problem in \eqref{Optimization_single_location} can be rewritten as \eqref{Optimization_single_location_simplified} on the top of next page, 
\begin{figure*}
     \begin{align}
        \max_{0\leq\ell_{m}\leq L_m}  2 
        \Re\set{ \ba_m^{\mathsf{T}}\tilde{\bg}_{m}\brc{\ell_m} -\bb_m^{\mathsf{T}}\tilde{\bg}_{m}\brc{\ell_m}} 
        -[\mW\mW^\her]_{m,m} \tilde{\bg}_{m}^{\her}\brc{\ell_m}\mU \tilde{\bg}_{m} \brc{\ell_m} ,\label{Optimization_single_location_simplified}
    \end{align}
    \hrule
\end{figure*}
    where $\ba_m$ is the $m$-th row of $\mW\mT^\her$, and 
    \begin{align}
    \bb_m^{\mathsf{T}}=\sum_{m'\ne m}[\mW\mW^\her]_{m,m'}\bh_{m'}^{\her}\mU.
    \end{align}
    By denoting $\bc_m =\ba_m -\bb_m$ and using the definition of $\tilde{\bg}_m \brc{\ell_m}$ given in \eqref{definition_h_l_m}, we can transform the objective function of the optimization \eqref{Optimization_single_location_simplified} into \eqref{eq:XI} given at the top of next page,
    \begin{figure*}
    \begin{align}
    &f_{m}(\ell_m) 
    =
        \sum_{k}
    \frac{ \xi \alpha_{m,k}\abs{c_{m,k}} }{ 
    D_{m,k}\brc{\ell_m}
    }
    \left[
    2 \cos\brc{ k_0\brc{
    D_{m,k}\brc{\ell_m} + i_{\mathrm{ref}} \ell_{m}}}-\angle{c_{m,k}}  -[\mW\mW^\her]_{m,m}
    { \xi \alpha_{m,k}
    \lambda_k\abs{q_k}^2
    }/{
    D_{m,k}\brc{\ell_m}
    }
    \right],\label{eq:XI}
    \end{align}
    \hrule
        \end{figure*}
    where $c_{m,k}$ is the $k$-th entry of $\mathbf{c}_m$, and $\angle{z}$ is the phase of $z$.
    
    The optimization of $\ell_m$ is hence carried 
    by classical scalar optimization, i.e., maximizing $f_{m}(\ell_m)$ in \eqref{eq:XI}, within a fixed interval, which can be effectively solved via grid search. 
    
\begin{remark}
It is worth mentioning that the classical gradient-based methods are not feasible for solving this scalar problem, since the objective function $f_m(\cdot)$ contains numerous stationary points caused by the oscillations of the cosine term.
\end{remark}

\begin{center}
\begin{minipage}{.95\linewidth}
\begin{algorithm}[H]
  \caption{\ac{fp}-\ac{bcd} Algorithm}
  \label{Algorithm1}
  \begin{algorithmic}[1]
  \STATE Choose a small  $\epsilon$ and initialize $\mathbf{W}$ and $\bl$ 
    \REPEAT
      \STATE Update $\omega_k$ and $q_k$ using \eqref{fp_bcd_omage_k} and \eqref{fp_bcd_q_k} for $k\in[K]$
      \STATE Update $\mathbf{W}$ according to \eqref{fp_bcd_W}
      \STATE Update $\ell_m$ by grid search for $m\in[M]$ 
    \UNTIL{increase of the objective in \eqref{eq:optim} falls below $\epsilon$}
    \STATE Scale $\mathbf{W}$ as ${\mW} =\sqrt{{P}/{\tr{{\mW}^{\her} {\mW}}}} {\mW}$
  \end{algorithmic}
\end{algorithm}
\end{minipage}
\end{center}

\subsection{Overall Algorithm, Convergence, and Complexity}
According to the solutions developed for each optimization block, the overall \ac{fp}-\ac{bcd} algorithm for solving problem \eqref{eq:optim} is given in Algorithm \ref{Algorithm1}. Since the objective value is non-decreasing in each step of block coordinate descent and is bounded from above due to the power constraint, the convergence of Algorithm \ref{Algorithm1} is guaranteed. Furthermore, Algorithm \ref{Algorithm1} is computationally efficient, as the optimization variables in each step are updated by either the closed-form solution or the low-complexity one-dimensional search. It can be shown that the complexities of updating $\omega_k$, $q_k$, $\mathbf{W}$, and $\ell_m$ in Algorithm \ref{Algorithm1} are in order of ${\mathcal{O}}(K^2M)$, ${\mathcal{O}}(K^2M)$, ${\mathcal{O}}(KM^2\!+\!M^3)$, and ${\mathcal{O}}(MLK)$, respectively. Hence, the overall complexity of Algorithm \ref{Algorithm1} scales with ${\mathcal{O}}(I(2K^2M\!+\!KM^2\!+\!M^3\!+\!MLK))$ with $I$ being the number of iteration for the algorithm convergence.

\begin{figure*}[!t]
\centering
    \subfigbottomskip=5pt
	\subfigcapskip=0pt
\setlength{\abovecaptionskip}{0pt}
    \subfigure[$K=M=4$ and $D=30$ m.]
    {
        \includegraphics[height=0.25\textwidth]{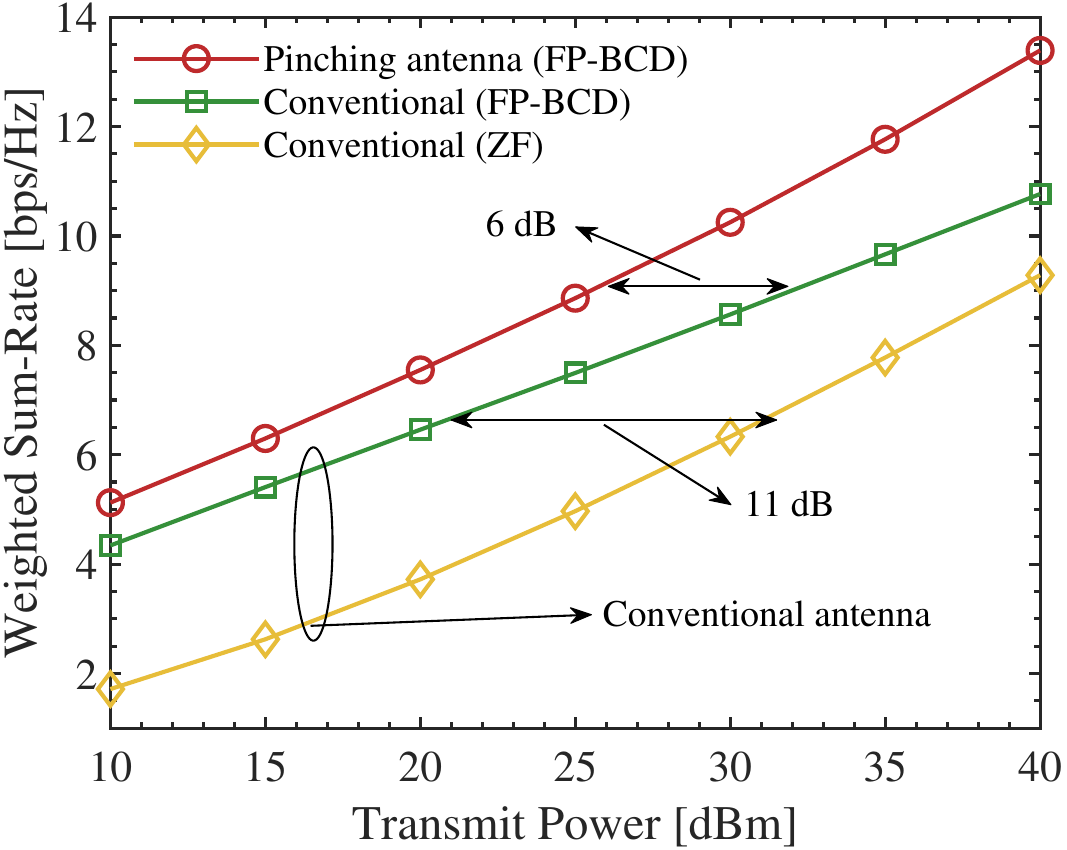}
	   \label{Figure: Sum_Rate_SNR}
    }
    \subfigure[$M=4$ and $P=20$ dBm.]
    {
        \includegraphics[height=0.25\textwidth]{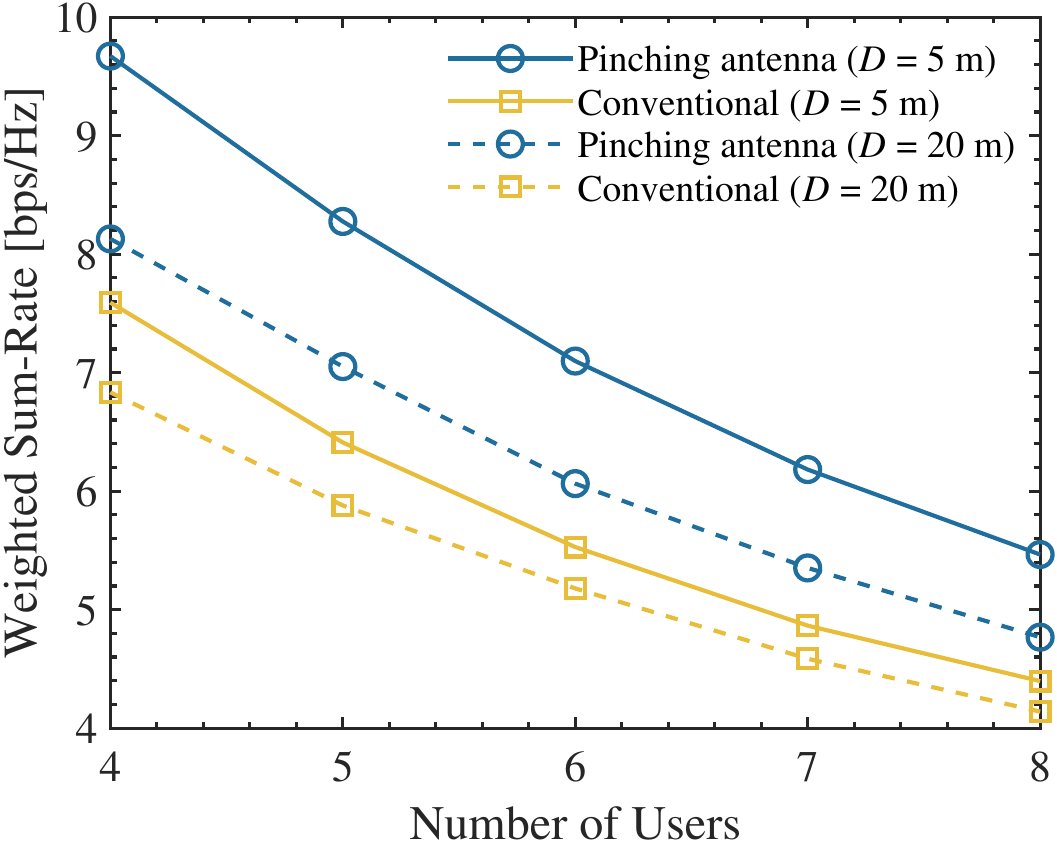}
	   \label{Figure: Sum_Rate_User}
    }
   \subfigure[$K=4$ and $P=20$ dBm.]
    {
        \includegraphics[height=0.25\textwidth]{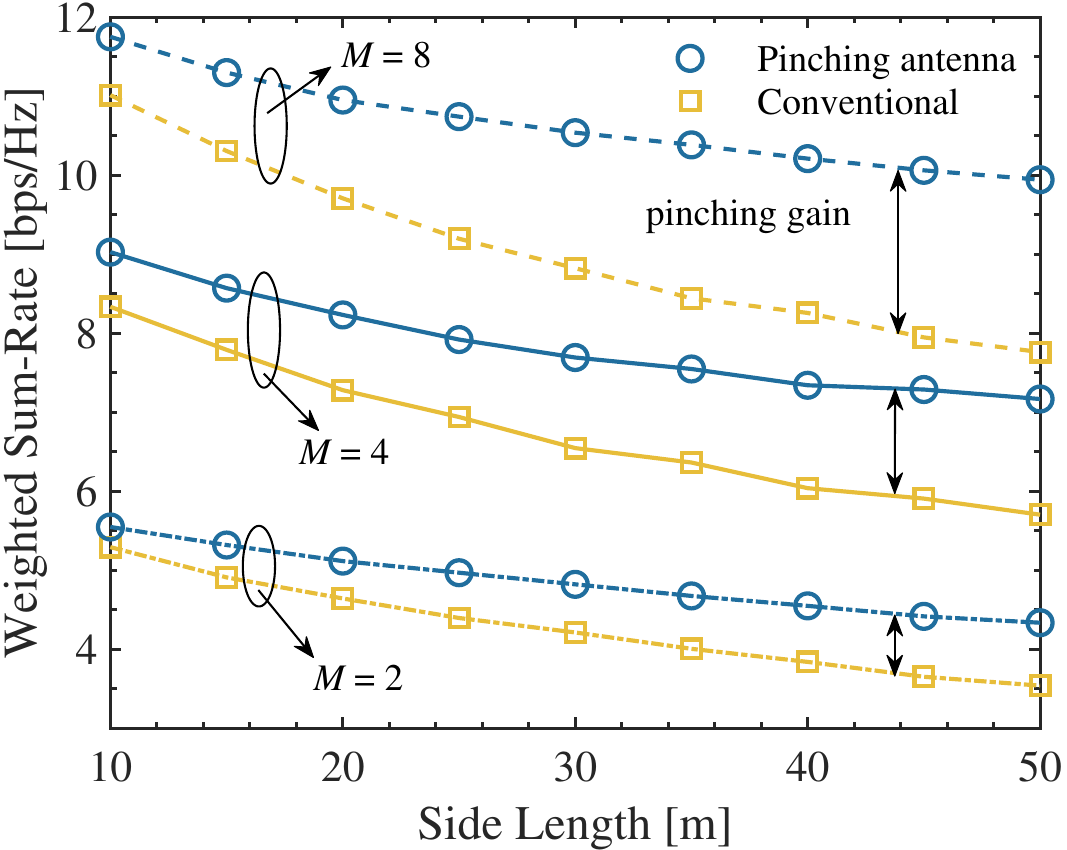}
	   \label{Figure: Sum_Rate_Range}
    }
\caption{Weighted sum-rate achieved by PAS. Arrows indicate the gain achieved by the proposed scheme over the fixed-antenna system.}
\vspace{-15pt}
\end{figure*}

\section{Numerical Investigations}
In this section, we validate the proposed algorithm through numerical experiments and compare it against the baseline.

\paragraph*{Experimental Setting}
The $K$ users are assumed to be randomly and uniformly distributed within a square region with side length $D$, centered at $\left[D/2,D/2,0\right]$, as illustrated in {\figurename} {\ref{Fig1_Schematic}}. Unless stated otherwise, the following parameters are used: the noise variance is set to $\sigma^2=-90$ dBm, the waveguides are located at height $a=3$ m and cover the entire area by being distanced as $d=\frac{D}{M-1}$ and of length ${\ell_m}\in\left[0,D\right]$, the carrier frequency is set to $f=28$ GHz and the reflective index is set to $i_{\mathrm{ref}}=1.44$. We assume uniform weighting, i.e., $\lambda_k=1/K$, and free-space path loss, i.e., $\alpha_{m,k}=1$ for $m\in[M]$ and $k\in[K]$. 

In the \ac{fp}-\ac{bcd} algorithm, the convergence threshold is set to $\epsilon=10^{-3}$. For the one-dimensional grid search, we set the number of points to $10^3$. Since the performance of the \ac{fp}-\ac{bcd} algorithm may depend on the initial parameters, the digital beamformer is initialized using the \ac{mrt} technique. The locations of the pinching elements are initialized based on a nearest-neighbor principle, where each $\ell_{m}$ is set to the location nearest to one of the users, i.e., $\ell_m=x_{k_m}$ for $m\in [M]$, where
\begin{align}
k_m = \argmin\nolimits_{k} \abs{y_k-(m-1)d}.
\end{align}
The numerical results are evaluated by averaging over $500$ random channel realizations. For performance comparison, the conventional antenna system serves as the baseline, where $M$ fixed-location antennas are deployed along the $y$-axis, centered within the square region. The antennas are half-wavelength spaced, with coordinates $[D/2,(m-1)\lambda/2,a]$ for $m\in[M]$.

\paragraph*{Numerical Results}
{\figurename} {\ref{Figure: Sum_Rate_SNR}} shows the weighted sum-rate as a function of the maximum total transmit power, $P$, which compares the performance of \ac{pas} with the conventional fixed-location antenna system using the \ac{fp}-\ac{bcd}-based and \ac{zf}-based beamforming schemes. The results indicate that the sum-rate increases with $P$ for all schemes. Notably, the \ac{pas} system can achieve significantly higher throughput than the conventional setting. To achieve the same sum-rate, the \ac{pas} requires considerably less transmit power. Specifically, a $6$ dB performance gain is observed for the \ac{pas} in comparison with the baseline when \ac{fp}-\ac{bcd} algorithm is employed. This gain grows to $11$ dB, when \ac{zf} beamforming is deployed at the conventional setting; see {\figurename} {\ref{Figure: Sum_Rate_SNR}}.

\begin{figure}[!t]
\centering
\setlength{\abovecaptionskip}{3pt}
\includegraphics[height=0.25\textwidth]{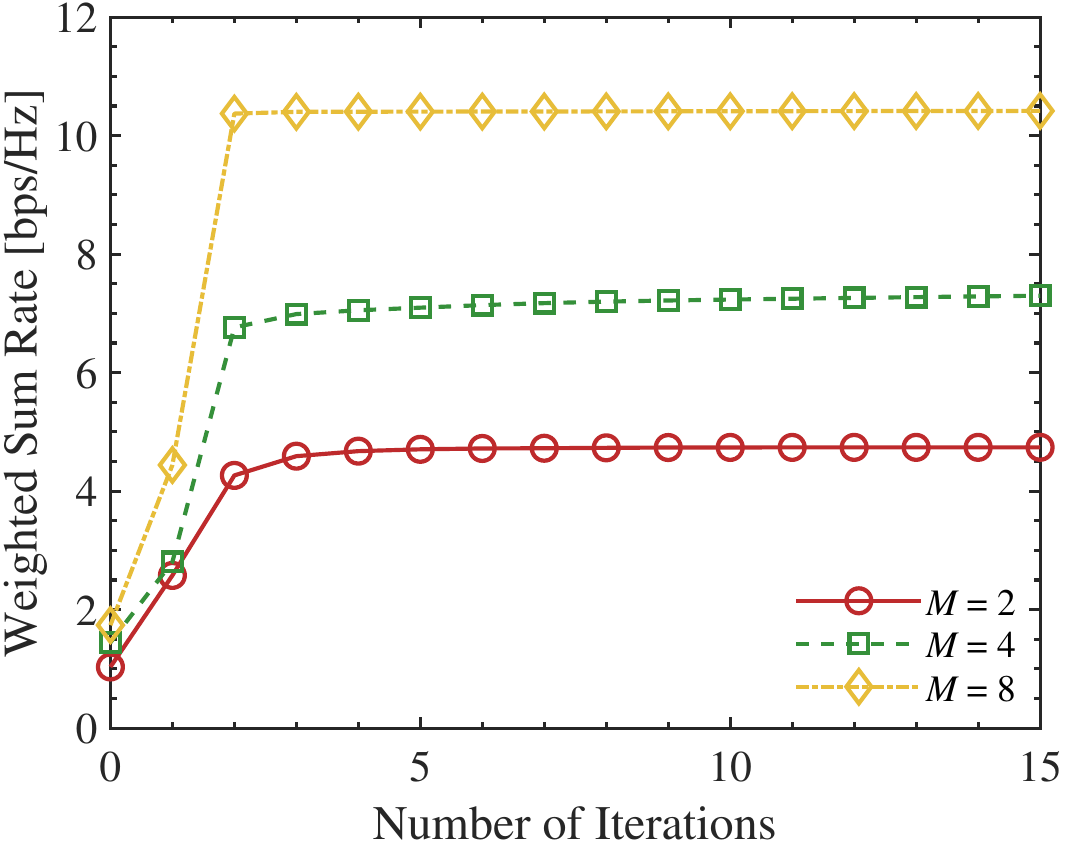}
\caption{Convergence of the proposed algorithm for different choices of $M$. Here, $P = 20$ dBm, $D=30$ m, and $K=4$.}
\label{Figure: Convergence}
\vspace{-15pt}
\end{figure}



{\figurename} {\ref{Figure: Sum_Rate_User}} depicts the sum-rate against the number of users $K$. The sum-rates achieved by both the proposed \ac{pas} setting and the baseline decrease monotonically as the number of users increases, primarily due to the rise in inter-user interference. {\figurename} {\ref{Figure: Sum_Rate_User}} also highlights that the performance gain of \ac{pas} over the baseline grows with the increase in the side length $D$, i.e., as the area size increases. This demonstrates the unique capability of \ac{pas} to establish strong \ac{los} links, as well as its capability to mitigate large-scale path-loss.


{\figurename} {\ref{Figure: Sum_Rate_Range}} further reinforces this observation by showing the sum-rate as a function of the side length of the square region. As the side length increases, the performance gain of the \ac{pas} over the conventional baseline improves significantly. This is due to the fact that a larger side-length increases the average distance between users and the center, which results in higher path loss for the conventional setting. In contrast, the \ac{pas} can flexibly position its elements close to the users to reduce the distance and enhance the throughput.

Finally, {\figurename} {\ref{Figure: Convergence}} illustrates the convergence behavior of the proposed \ac{fp}-\ac{bcd} algorithm for different numbers of antenna elements, $M$. As observed, the weighted sum-rate gradually increases with more iterations and finally converges to a stable value for all considered values of $M$, which confirms the effectiveness of the proposed algorithm. 

\section{Conclusion}
We studied downlink beamforming in a multiuser \ac{mimo} system, which deploys an array of pinched waveguides for transmission. A low-complexity joint beamforming design for this \ac{pas}-aided \ac{mimo} was developed, which invokes the principles of \ac{fp} and \ac{bcd} to alternatively optimize the digital beamformer at the \ac{ap} and the activated locations of the pinching elements on the waveguides, aiming to maximize the achievable sum-rate. The effectiveness of the proposed design was validated through several numerical experiments. The results of this study showcase the superiority of \ac{pas}-aided \ac{mimo} over conventional fixed-location designs, which together with its practicality highlight the potentials of \ac{pas}.
\bibliographystyle{IEEEtran}
\bibliography{ref}
\end{document}